\documentstyle[12pt]{article}
\def\({\c c}
\def\|{\'\i}

\def\({\c c}
\def\|{\'\i}

\begin{document}

\baselineskip=15pt   

\begin{flushright}
SLAC-PUB-8016\\
November 1998
\end{flushright}

\bigskip
\bigskip
\baselineskip=18pt  %

\begin{center}
{\large\bf  
LIGHT-FRONT QUANTIZED CHIRAL SCHWINGER MODEL AND ITS VACUUM STRUCTURE}
\footnote{Research partially supported
by the Department of Energy under contract DE-AC03-76SF00515
}

\vspace{1.0cm}

\baselineskip=16pt  %
{\large Prem P. Srivastava}\footnote{E-mail:\quad prem@slac.stanford.edu 
or prem@lafexsu1.lafex.cbpf.br. On
leave of absence from {\it Instituto de F\'{\i}sica, UERJ-Universidade do 
Estado de Rio de Janeiro, Brasil}.}

\vspace{0.4cm}
{\em \it Stanford Linear Accelerator Center, Stanford University, 
Stanford, California 94309}
\vspace{0.7cm}

{\bf Abstract}

\end{center}

\vspace{0.2cm}
\baselineskip=14pt
 {\small The bosonized Chiral 
Schwinger model (CSM) is quantized 
on the light-front (LF). The physical 
Hilbert space of CSM is obtained directly once the constraints 
on the LF phase space are 
eliminated. The discussion of the degenerate vacua and the 
absence in the CSM of the  $\theta$-vacua, as found in 
the Schwinger model (SM), becomes straightforward. 
The differences in the structures of the the mass excitations 
and the vacua in these gauge theories are displayed  
transparently. The procedure  
followed is the one used successfully in the previous works  
for describing  the spontaneous symmetry breaking (SSB) 
and the SM on the LF.  The physical contents following from
the  LF quantized theory agree with those known in the 
conventional treatment. The LF hyperplane is argued to be equally
appropriate as the conventional equal-time one for the canonical 
quantization. Some comments on the irrelevance, in quantized 
field theory, of the fact that the hyperplanes $x^{\pm}=0$ 
constitute characteristic surfaces of hyperbolic partial 
differential equation are also made.} 


\newpage

\setcounter{footnote}{0}         %

\baselineskip=16pt   %

\section{Introduction}\label{intro}

Dirac \cite{dir}, in his paper in 1949, discussed the problem 
of constructing a dynamical theory of physical system which would 
incorporate  
in it the principle of quantization together with that of 
the special relativity theory. The 
LF quantization which studies the  relativistic 
quantum dynamics on the hyperplanes  
: $x^{0}+x^{3}\equiv {\sqrt{2}}x^{+}=const.$,  called 
the {\it front form} theory, was also proposed there. 
 The {\it instant form} or the 
conventional equal-time theory 
on the contrary uses the $x^{0}=const.$ hyperplanes. The former studies the 
evolution of the relativistic dynamical system in $x^{+}$ 
while the latter in $x^{0}$.    
The LF coordinates $x^{\mu}: (x^{+},x^{-},x^{\perp} 
)$,  where $x^{\pm}=(x^{0}{\pm} x^{3}) 
/{\sqrt 2}=x_{\mp}$ and   $ x^{\perp} = 
(x^{1}, x^{2})$,   are convenient to use in the {\it front form}
theory.  
They are  {\it not related by a Lorentz transformation} 
to the coordinates $(x^{0}\equiv t,x^{1},x^{2},x^{3})$ 
usually employed in the {\it instant form } theory and as such the 
same physical content in a dynamical theory may acquire  
different description in the two treatments.  The discussion from the 
LF quantized field theory may also   
 be of relevance towards the understanding, say, 
of the simultaneous inclusion in dynamical theory of the 
principles of the general covariance and the quantization\footnote{
We recall the experience with the discovery of 
the Kruskal-Szekers coordinates in early sixtees 
which shed a new light on the problem of the Schwarzshild singularity in the
theory of  gravitation.}.

We will make the {\it convention} to regard\footnote{
The coordinates $x^{+}$ and $x^{-}$ appear in a symmetric fashion 
and we note that $\left[x^{+},{1\over i}\partial^{-} \right]
= \left[x^{-},{1\over i}\partial^{+} \right]=i $ where $\partial^{\pm}=
\partial_{\mp}=(\partial^{0}\pm \partial^{3})/\sqrt {2}$ etc.. }
$x^{+}$ as the 
LF-time coordinate while $x^{-}\equiv x$ as the {\sl longitudinal 
spatial} coordinate. The (temporal)  evolution in $t$ or 
$x^{+}\equiv \tau$ of the system is 
generated by Hamiltonians which are very different 
in the two {\it forms} of the theory. 

Consider \cite{pre} the invariant distance between two spacetime points 
: $ (x-y)^{2}=(x^{0}-y^{0})^{2}-(\vec x-\vec y)^2= 2 
(x^{+}-y^{+}) (x^{-}-y^{-}) - (x^{\perp}-y^{\perp})^{2}$. 
On an equal $x^{0}=y^{0}=const. $ hyperplane the points have  
spacelike separation  except for if  they 
are {\it coincident} when it becomes lightlike one.  
On the LF with $x^{+}=y^{+}=const.$ 
the distance becomes  {\it independent of}  $(x^{-}-y^{-})$ and 
the seperation is again spacelike; it becomes lightlike one 
when  $x^{\perp}=y^{\perp}$ but with the difference that 
now the points need {\it not}  
necessarily be coincident along the longitudinal direction. 
The LF field theory hence need not necessarily be local 
in $x^{-}$, even if the corresponding 
{\it instant form} theory is given to be a local one in all the three 
spatial coordinates $\vec x$.  
For example, the commutator 
$[A(x^{+},x^{-},{x^{\perp}}),B(0,0,0^{\perp})]_{x^{+}=0}$ 
of two scalar observables would vanish on the grounds of
microcausality principle if   
$ x^{\perp}\ne 
0$ since   $x^{2}\vert_{x^{+}=0}$ is spacelike. 
Its value  would  be thus proportional to  $\,\delta^{2}(x^{\perp})\, $ 
and a finite number of its derivatives,  
implying locality only in $x^{\perp}$ but not necessarily so 
in $x^{-}$. Similar arguments in 
the {\it instant form} theory lead to the locality 
in all the three spatial coordinates. 
Both  of the commutators 
 $[A(x),B(0)]_{x^{+}=0}$ and 
$[A(x),B(0)]_{x^{0}=0}$ are nonvanishing    
only on the light-cone. 

We remark that in the LF 
quantization we time order with 
respect to   $\tau$ rather than  $t$.  
The  microcausality principle, 
however, ensures that the retarded commutators  
$[A(x),B(0)]\theta(x^{0})$ and  $[A(x),B(0)]\theta(x^{+})$,  
which appear \cite{ryd} in the S-matrix elements,  
do not  lead to disagreements in the two formulations. 
In the regions 
$x^{0}>0, x^{+}<0$ and $x^{0}<0, x^{+}>0$, where the commutators  
seem different  the $x^{2}$  is spacelike. 
Hence, if we assume the {\it microcausality
principle},  the LF hyperplane  
seems  {\it equally appropriate} as the conventional  one 
of the {\it instant form} theory for the canonical quantization.

The structure of the  phase space in the {\it front form} theory is 
different from that of the one in the  conventional theory.     
For example, the LF vacuum 
is generally  found simpler \cite{bro, ken} and in many cases 
the interacting theory vacuum 
is seen to coincide  with the perturbation theory one. 
The SSB in the scalar theory  is also   
described \cite{pre} differently on the LF. 
The broken continuous symmetry is  inferred  now 
from the residual symmetry of the LF Hamiltonian 
operator while   the symmetry of the LF vacuum remains 
unbroken, which is  in contrast to the conventional 
description in which the 
symmetry of the vacuum state is broken while the Hamiltonian remains 
invariant.  
The expression which counts the number of Goldstone bosons in the {\it 
front form} theory, however, is 
found to be the same as 
in the conventional treatment. The Coleman's theorem on the absence 
of the Goldstone bosons in two dimensional scalar theory also finds a
new demonstration \cite{pre} in the {\it front form} theory. 

 A recent study \cite{pre1} on the 
LF quantized SM showed that we are led directly to the 
{\it physical Hilbert space}  once the 
constraints on the phase space are eliminated. The 
well known \cite{low} {\it condensate} or 
$\theta$-vacua and their continuum normalization  
were  shown to emerge \cite {pre1} in a straightforward fashion. 
In the present work we study  the bosonized CSM on the LF and 
demonstrate in 
equally direct fashion  its degenerate vacuum structure along with 
the {\it absence of the condensate or $\theta$-vacua} in this model.

An important advantge pointed out by Dirac of the {\it front form} 
theory is that in it  {\it seven} out of the ten   
Poincar\'e generators are {\it kinematical}, e.g., they leave 
the hyperplane $x^{+}=0 $ invariant \cite{dir}. They are\footnote{ 
In the standard notation $K_{i}=-M^{0i}, J_{i}=-(1/2)\epsilon_{ijk}
M^{kl}, i,j,k=1,2,3$. The generator 
$K_{3}$ is dynamical one in the {\it instant form} theory. It  
is in contrast kinematical in the {\it front form} theory where it 
generates  the scale 
transformations of the LF components of $x^{\mu}$, 
$P^{\mu}$ and  $M^{\mu\nu}$, with  $\mu,\nu=+,-,1,2$ and where $P^{\pm}=
(P^{0}\pm P^{3})/\sqrt{2}$ etc..  }
$P^{+},P^{1},P^{2},\, 
M^{12}=-J_{3},\, M^{+-}= M^{03}= -K_{3},\,
M^{1+}=(K_{1}+J_{2})/\sqrt{2}$ and 
$ M^{+2}=(K_{2}-J_{1})/\sqrt{2}$. 
In the conventional theory  
 only six such ones \cite{dir}, {\it viz.}, ${\vec P}$ and  
$M^{ij}=-M^{ij} $,  leave the hyperplane $x^{0}=0$ 
invariant.

We recall also that the 
LF field theory was rediscovered \cite{wei} by Weinberg  in his 
Feynman rules adapted for the infinite momentum frame. It  was 
demonstrated \cite{kog} latter that these  rules, in fact, correspond to 
the {\it front form} quantized theory.  
It was also  successfully employed    in the {\it nonabelian bosonization
} of the field theory of N free Majorana fermions,  
where the corresponding LF current algebra  was compared \cite{wit} 
with the one in the bosonized theory described by the WZNW action 
at the critical point. 

 The interest in LF quantization  
has been revived \cite{bro, ken} also  
due to the difficulties encountered 
in the computation, in the conventional 
framework, 
 of the nonperturbative effects in the context of  QCD 
and  the problem of the relativistic bound states of light 
fermions \cite{ken, bro} 
in the presence of the complex vacuum structure.  
  The {\it front-form} dynamics may serve as a 
complementary tool where we have a simple vacuum while the 
complexity of the problem is now transferred to the 
LF Hamiltonian. 
In the case of the scalar field theory, for example, 
the LF  Hamiltonian is in fact
found \cite {pre2, pre}  
to be nonlocal due to the presence of\footnote{In fact, Dirac \cite{dir} 
in his paper does give an example showing that the potential must
 be constrained if we incorporate in the dynamical theory the principles 
of quantization and special relativity.}     
{\it constraint equations} in the Hamiltonian formulation.

The chiral $QED_{2}$ or CSM, employing the conventional 
framework, has received \cite{abd,
bas} much attention  since 
Jackiw and Rajaraman \cite{jac} pointed out that, despite the gauge 
anomaly it developed due to the renormalization 
ambiguity, the theory can be shown to be unitary and consistently
quantized.  

The procedure used \cite{pre, pre2} 
previously for explaining  the SSB on the LF and recently 
\cite{pre1} in the  bosonized SM is applied below to discuss the CSM. 
The  scalar field is first 
separated, based on physical considerations, into the {\it dynamical 
bosonic condensate} variable $\omega(\tau,x^{\perp})$ 
and the quantum fluctuation field $\varphi(\tau,x^{-},x^{\perp})$, 
e.g.,  $\phi(\tau,x^{-},x^{\perp})= \omega (\tau,x^{\perp})+\varphi(
\tau,x^{-},x^{\perp})$. The {\sl standard} Dirac method \cite{dir1} 
is subsequently 
applied to construct the self-consistent 
LF Hamiltonian framework which is  then 
 quantized canonically. The c- or q-number nature of the condensate 
$\omega$ emerges from  inside the theory itself.

Sec. 2 discusses how the condensate variable is subtracted out by
simple field redefinition from the Lagrangian of the 
 bosonized CSM on the LF. 
The canonical Hamiltonian framework 
is constructed in Sec. 3 following the standard Dirac method. 
Its quantization, 
the structure of the Hilbert space, the degenerate vacua and the 
mass spectrum are 
studied in Sec. 4. Conclusions are summarized in Sec. 5 where 
some comments are also made  on the relevance to the LF quantization 
of the fact that  
the   $x^{\pm}=0$ hyperplanes are the  characteristic surfaces of
hyperbolic partial differential equation. In order to solve 
the Cauchy initial value problem in the classical theory of 
partial differential equations we would be required to specify the data on 
both of these surfaces; in the context of the LF quantization we need 
to select  only one of the hyperplanes.

\begin{sloppypar}

\section{Bosonized CSM on the LF. Absence of $\theta$-vacua}\label{bosmod}
The  Lagrangian density of the chiral $QED_{2}$ or CSM  model under 
consideration is

\begin{equation}
{\cal L}= -{1\over 4}F^{\mu\nu}F_{\mu\nu} 
+  {\bar\psi}_{R}\,i\gamma^{\mu}\partial_{\mu}
\psi_{R}+ {\bar\psi}_{L}\,\gamma^{\mu}(i\partial_{\mu}+2e\sqrt{\pi}
 A_{\mu})\psi_{L},
 \end{equation}
where\footnote{ Here 
$\gamma^{0}=\sigma_{1}$, $ \gamma^{1}=i\sigma_{2}$, 
$\gamma_{5}=-\sigma_{3}$, 
  $x^{\mu}:\,(x^{+}\equiv \tau,x^{-}\equiv x)$ 
with ${\sqrt 2}x^{\pm}={\sqrt 2}x_{\mp}=(x^{0}{\pm} x^{1})$, 
$A^{\pm}=A_{\mp}=(A^{0}\pm A^{1})/
{\sqrt 2}$,  
$\psi_{L,R}=P_{L,R}\;\psi $, $P_{L}=(1-\gamma_{5})/2$, 
$P_{R}=(1+\gamma_{5})/2$, 
$\bar\psi= \psi^{\dag}\gamma^{0}$.} 
  $\psi= \psi_{R}+\psi_{L}$ is a two-component spinor field and
$A_{\mu}$ is the abelian gauge field
The classical Lagrangian (1) is  
 invariant under the local $U(1)$ gauge transformations $A_{\mu}\to 
A_{\mu}+\partial_{\mu}\alpha/(2\sqrt{\pi}e)$, $\psi\to [P_{R}+
e^{i\alpha}P_{L}] \psi $ and under the global 
$U(1)_{5}$ chiral transformations $\psi\to exp(i\gamma_{5}
\alpha)\,\psi $.

The  model under study can be solved completely using the technique of
bosonization. The latter consists in the replacement of a known
system of fermions with a theory of bosons which has a completely
equivalent physical content, including, for example, identical
spectra, the same current commutation relations and the
energy-momentum tensor when expressed in terms of the currents. The 
bosonized  version of (1) is convenient to study the vacuum structure
and it was shown \cite{jac} to be   
\begin{equation} 
    S = \int d^2x \left[ -{{1}\over {4}}F_{\mu \nu}F^{\mu \nu} 
                          +{{1}\over{2}}\partial_{\mu}\phi\partial^{\mu}\phi
                          +eA_{\nu}(\eta^{\mu \nu}
                          -\epsilon^{\mu \nu})\partial_{\mu}\phi
              +{{1}\over{2}}ae^{2}A_{\mu}A^{\mu}\right] 
\end{equation}
Here the explicit mass term for the gauge field 
parametrized by the constant parameter $a$ represents a 
regularization ambiguity 
and the breakdown of $U(1)$ gauge symmetry. The action (2) may be
derived by the functional integral or the canonical
quantization methods.

Following the procedure successfully used in the earlier works we first make 
  the {\it separation}:     
$\,\phi(\tau,x^{-})= \omega (\tau)+\varphi(
\tau,x^{-})$. The subsequent application of the Dirac method then 
enabled us to give \cite{pre, pre2} the  
description  on the LF of the SSB in the scalar theory and also the variable 
 $\omega$ was shown there 
to come out as a c-number (background field).   
On the other hand in the bosonized SM on the LF it turned out 
to be  q-number operator whose eigenvalues were  shown \cite{pre1}  
to label the {\sl condensate} or $\theta$-vacua. 
We  set $\int dx^{-}\varphi(\tau,x^{-})=0$ so that the 
entire zero-momentum mode of 
$\phi$ is represented by the condensate variable and 
recall \cite{pre1} also that  the {\it chiral transformation} 
is defined by:\quad $\omega\to \omega+const., \; \varphi\to \varphi$,  
and $A_{\mu}\to A_{\mu}$. This ensures  
that the {\it boundary conditions} on the 
$\varphi$ are kept   
unaltered under such transformations and our mathematical framework 
may be considered {\it well posed}, before we proceed to build 
the canonical Hamiltonian framework. 

Written explicitly 
(2) takes the following form on the LF 
\begin{equation} 
 S = \int\, d^2x \;\left[{\dot\varphi}\varphi'+
{1\over 2}({\dot A}_{-}-{A'_{+}})^{2}+ a e^{2}[A_{+}+{2\over {ae}}(
\dot\omega+\dot\varphi)] A_{-} \right] 
\end{equation}
where an overdot (a prime) indicates the partial derivative 
with respect to $\tau$ ( $x$).  
In order to suppress the finite volume effects we    
work in the {\it continuum formulation}  and   
require, based  on physical considerations,   that the fields satisfy the 
boundary conditions needed 
for the existence of their Fourier transforms 
in the  spatial variable $x^{-}$. 
We note now that  $A_{+}$ appears in the action (3) 
as an {\it auxiliary} field, 
without a kinetic term. It is clear that 
the condensate variable may thus be subtracted out 
from the theory using the frequently adopted 
procedure of  
{\it field redefinition} \cite{nie} on it:  
$\;\;A_{+}\to A_{+}-2\dot\omega/(ae)$, obtaining thereby 

\begin{equation}
{\cal L}_{CSM}={\dot\varphi}\varphi'+
{1\over 2}({\dot A}_{-}-{A'_{+}})^{2}+ 2e {\dot\varphi} A_{-}+
a e^{2} A_{+} A_{-},   
\end{equation}
which signals the emergence of a  {\it different structure} of the
Hilbert space compared to that of the SM.  There\footnote{
In the SM we have \cite{pre1}:  
$ L= \int\, dx^{-}\;\Bigl[{\dot\varphi}\varphi'+{1\over 2}
({\dot A}_{-}-{A'_{+}})^{2}-(e/\sqrt {\pi})
(A_{+}\varphi'-A_{-}{\dot\varphi})\Bigr]+ 
(e/{\sqrt \pi}){\dot\omega}h(\tau) $  where 
$ h(\tau)=\int dx^{-} A_{-}(\tau,x^{-})$.} the  {\it condensate} or 
$\theta$-vacua emerged due to the presence of the  additional 
variable $\omega$ in the theory. 

\section{LF Hamiltonian Framework}\label{hamilton}
The Lagrange eqs. following from (4) are 

\begin{eqnarray}
    \partial_{+}\partial_{-}\varphi~&=&~-e \partial_{+}A_{-}, \nonumber \\
    \partial_{+}\partial_{+}A_{-}-\partial_{+}\partial_{-}A_{+} ~&=&~
a e^{2} A_{+} +2 e \partial_{+}\varphi, \nonumber \\
\partial_{-}\partial_{-}A_{+}-\partial_{+}\partial_{-}A_{-}
  ~&=&~ a e^{2} A_{-}. 
\end{eqnarray}
and for $ a\neq 1 $ they lead to: 
\begin{eqnarray}
       \Box G(\tau,x)& = & 0 \nonumber\\
\left[ \Box + {e^2a^2\over (a-1) }\right] E(\tau,x) & =& 0, 
\end{eqnarray}  
where $E=(\partial_{+}A_{-}-\partial_{-}A_{+})$ and 
$G= (E-ae\varphi)$. 
Both the massive and massless scalar  excitations 
are present in the theory and  the tachyons
would be absent in the specrtum if   $a>1$; the case  considered 
in this paper. 
 We will confirm in the Hamiltonian framework below 
that the $E$ and $G$ represent, in fact, the two 
independent field operators on the  LF phase space. 

The Dirac procedure \cite{dir1} as applied to (4) 
is straightforward. The canonical momenta  are 
$    \pi^{+}\approx 0,     \pi^{-}\equiv E=
 \dot{A}_{-}- A'_{+},   \pi_{\varphi}= {\varphi}'+2e A_{-}$ which
result in 
two primary  weak constraints  
$    \pi^{+} \approx 0 $ and 
$ \Omega_1 \equiv (\pi_{\varphi}-\varphi'-2eA_{-})\approx 0$. A 
secondary constraint 
$    \Omega_2 \equiv \partial_{-} E + 
a e^{2} A_{-} \approx 0$ is shown to emerge 
when we require the $\tau$ independence 
(persistency) of 
$\pi^{+}\approx 0$ employing the preliminary  Hamiltonian
\begin{equation}
    H' = {H_c}^{lf} + \int dx~u_{+}\pi^{+}+\int dx~u_{1}\Omega_1 ,
\end{equation}
where $u_{+}$ and $u_{1}$ are the Lagrange multiplier fields and 
${H_c}^{lf}$ is the canonical Hamiltonian  
\begin{equation}
    {H_c}^{lf} = \int\; dx~\left[~
            \frac{1}{2}{E}^{2} + E A_{+}'
-ae^{2} A_{+}A_{-}
               \right]. 
\end{equation}
and we assume initially the  standard   
equal-$\tau$ Poisson  
brackets : 
 $\{E^{\mu}(\tau,x^{-}),A_{\nu}(\tau,
y^{-}) \}=-\delta_{\nu}^{\mu}\delta (x^{-}-y^{-})$,  
$\{\pi_{\varphi}(\tau,x^{-}),\varphi(\tau,y^{-}) \}
=-\delta(x^{-}-y^{-})$ etc.. 
 The persistency requirement for $\Omega_{1} $ results in an equation for 
determining  $u_{1}$. The procedure is repeated with the 
following extended 
Hamiltonian which includes in it also   the secondary constraint  
\begin{equation}
    {H_e}^{lf} = {H_c}^{lf} + \int dx~u_{+}\pi^{+}+\int dx~u_{1}\Omega_1 
+\int dx~u_{2}\Omega_{2}. 
\end{equation}
No more secondary constraints are seen 
to arise; we are left with the persistency conditions which
determine the multiplier fields  $u_{1}$ and $u_{2}$ while $u_{+}$
remains undetermined. We also  
find\footnote{\quad We make the convention that the first variable in an equal-
$\tau$ bracket refers to the longitudinal coordinate $x^{-}\equiv x$
while the second one to $y^{-}\equiv y$ while  $\tau$ is suppressed.} 
 $(C)_{ij}=\{\Omega_{i},\Omega_{j}\}~=~D_{ij}~$ $ (-2
\partial_{x} \delta(x-y))$ where $i,j =1,2$ and $~D_{11}=1,~D_{22}=a
e^2, ~D_{12}=~D_{21}= -e$ and  that $\pi^{+}$ has 
vanishing brackets with $\Omega_{1,2}$. 
The $\pi^{+}\approx 0 $ is     first class weak 
constraint while 
 $\Omega_1$ and $\Omega_2 $, which does not depend on  $A_{+}$ or 
$\pi^{+}$,  are second class ones. 
\end{sloppypar}

\begin{sloppypar}

We  go over from the Poisson bracket  
to the  Dirac bracket $\{,\}_{D}$ 
 constructed in relation to the pair,    
 $\Omega_1\approx 0$ and   $\Omega_2\approx 0 $ 

\begin{equation} 
\{f(x),g(y)\}_{D}=\{f(x),g(y)\}-\int\int du dv\;\{f(x),\Omega_{i}(u)\}
(C^{-1}(u,v))_{ij}\{\Omega_{j}(v), g(y)\}. 
\end{equation}
Here $C^{-1}$ is the inverse of $C$ and we find 
$(C^{-1}(x,y))_{ij}=B_{ij}$ $K(x,y)$ with  $~B_{11}=a/(a-1)$,$
~B_{22}=1/[(a-1)e^2]$, $~B_{12}=B_{21}$$= 1/[(a-1)e],$ and 
$K(x,y)=-\epsilon(x-y)/4$.  Some of the Dirac brackets are 
$\{\varphi,\varphi\}_{D}= B_{11} ~K(x,y); 
~\{\varphi,E\}_{D} = e B_{11} ~K(x,y); 
~\{E,E\}_{D} = ae^{2} B_{11} ~K(x,y); 
~\{\varphi, A_{-}\}_{D}=-B_{12}~\delta(x-y)/2; 
~\{A_{-},E\}_{D}=B_{11}~\delta (x-y)/2;
~\{A_{-},A_{-}\}_{D}=B_{12}\partial_{x}~\delta(x-y)/2$  
and the only nonvanishing one involving $A_{+}$ or $\pi^{+}$ is 
$\{A_{+},\pi^{+}\}_{D}= \delta(x-y)$. 

\end{sloppypar}

The eqns. of motion employ now the Dirac brackets and 
inside them, in view of their very construction \cite{dir1}, we may set 
$\Omega_1=0$ and $\Omega_2=0 $  as strong relations. The 
Hamiltonian is therefore  effectively given by $H_{e}$ 
with the terms involving the multipliers $u_{1}$ and $u_{2}$ 
dropped. The multiplier $u_{+}$ is not determined since the constraint 
$\pi^{+}\approx 0$ 
continues to be first class even when the above 
Dirac bracket is employed. 
The variables $\pi_{\varphi}$ and $A_{-}$ are then removed from the 
theory leaving behind $\varphi$, $E$, $A_{+}$, and $\pi^{+}$  as 
the remaining independent variables. 
The canonical Hamiltonian density reduces to 
${\cal H}_{c}^{lf}= E^2/2 +\partial_{-}(A_{+}E)$ while $\dot A_{+}=
\{A_{+}, H_{e}^{lf}\}_{D}=u_{+}$. The surface term in 
the canonical LF Hamiltonian may be ignored if, 
say, $E (=F_{+-})$ vanishes at 
infinity. The variables $\pi^{+}$ and $A_{+}$ are then seen to describe 
a decoupled (from $\varphi$ and $E$) free theory 
and we may hence drop these variables as well. 
The effective LF  Hamiltonian thus takes the simple form 

\begin{equation}
H_{CSM}^{lf} = {1\over {2}} \int dx \; E^{2}, 
\end{equation}
which is to be contrasted with the one found 
in the conventional treatment \cite{bas, abd}. 
 $E$ and $G$ (or $E$ and $\varphi$) are now the independent variables
on the phase space and the eqs. (6) are verified to be
recovered for them which assures us of the selfconsistency
\cite{dir1}. 
We  stress that in our 
discussion we do {\it not} employ any gauge-fixing.  The same result
(11) could be alternatively  obtained\footnote{ A similar discussion is
encountered also in the LF quantized 
Chern-Simons-Higgs system \cite{pre3}.}, 
however,  
if we did introduce  the gauge-fixing constraint $A_{+}\approx 0$ 
and made further 
modification on $\{,\}_{D}$ in order to implement  $A_{+}\approx 0, 
\pi^{+}\approx0$ as well.  
That it is accessible  on the phase space to take care of the 
remaining first class constraint, but not in the Lagrangian in (4), 
follows from the Hamiltons 
eqns. of motion. We recall \cite{pre1} that in the SM
 $\varphi$, $\omega$, and $\pi_{\omega}=
(e/\sqrt {\pi})\int dx A_{-}$ were shown to be the 
independent operators 
 and that the matter field $\varphi$ appeared instead in the LF 
Hamiltonian.
\section{Quantization. Vacuum structure in CSM}\label{quant}
\begin{sloppypar}
The  canonical quantization is peformed 
via the correspondence $i\{f,g\}_{D}\to [f,g] $ and we find the
following equal-$\tau$ commutators 
\begin{eqnarray}
\left[ E(x),E(y) \right]&=& i K(x,y){ a^2 e^2/ (a-1)},\nonumber\\ 
\left[ G(x),E(y)\right]& =& 0,   \nonumber\\
\left[ G(x),G(y)\right]& =& {ia^2 e^2} K(x,y).
\end{eqnarray} 
For $a>1$, when the tachyons are absent as seen from (6), 
these commutators  are also 
physical and  the independent field operators $E$ and $G$  generate 
the Hilbert space with a tensor product structure 
of the Fock spaces 
$F_{E}$ and $F_{G}$ of these fields
 with  the positive definite metric. 

We can make, in view of (12),  the following LF 
momentum space expansions 
\begin{eqnarray}
E(x,\tau)&=&{ae\over {{\sqrt{(a-1)}}{\sqrt{2\pi}}}} \int_{-\infty}^{\infty} dk
\;{{\theta(k)}\over{\sqrt{2k}}}
\left[d(k,\tau)e^{-ikx}~+~d^{\dag}(k,\tau)e^{ikx}\right],\nonumber\\ 
G(x,\tau) &=& {{ae}\over {\sqrt{2\pi}}} 
\int_{-\infty}^{\infty} dk\;{{\theta(k)}\over{\sqrt{2k}}}
\left[g(k,\tau)e^{-ikx}~+~g^{\dag}(k,\tau)e^{ikx}\right], 
\end{eqnarray}
where the operators ($d, g, d^{\dag},g^{\dag}$) satisfy the 
canonical commutation relations of two independent harmonic
oscillators;   
the well known set of Schwinger's bosonic oscillators, often employed
in the angular momentum theory. The
expression for the Hamiltonian becomes 
\begin{equation}
H_{CSM}^{lf}= \delta(0){{a^2e^2}\over {2(a-1)}}~\int_{k>0}^\infty
{{dk}\over {2k}}
\; N_{d}(k,\tau)
\end{equation}
where we have dropped the infinite zero-point energy term and 
note that \cite{ryd} $\left[d^{\dag}(k,\tau),d(l,\tau)\right]=
-\delta(k-l)$, $d^{\dag}(k,\tau)d(k,\tau)= 
\delta(0) N_{d}(k,\tau)$ etc. 
 with similar expressions 
for the independent g-oscillators. 
We verify that $\left [N_{d}(k,\tau),N_{d}(l,\tau)\right]=0$, 
$\left [N_{d}(k,\tau),N_{g}(l,\tau)\right]=0$,  
$\left [N_{d}(k,\tau),d^{\dag}(k,\tau)\right]=  d^{\dag}(k,\tau) $
etc.. 
\end{sloppypar}

The Fock
space can hence be built on a basis of eigenstates of the 
hermitian number operators $N_{d}$ and  $N_{g}$. 
The   ground state of CSM is  degenerate and 
described by $\vert 0> = \vert E=0)\otimes \vert G\}$ and it carries 
vanishing LF energy. 
For a fixed $k$ these states,  $ \vert E=0)\otimes 
{({g^{\dag}(k,\tau)}^{n}/\sqrt {n!})}\vert 0\}$,  
are labelled by the integers $n=0,1,2,\cdots $. The $\theta$-vacua 
are absent in the CSM, however, we 
recall \cite {pre1} 
that in the SM the degenerate {\it chiral vacua} are also 
labelled  by 
such integers. We remark also that on the LF we work in the Minkowski 
space and that in our discussion we do {\it not} make use of the 
Euclidean space theory action, where the (classical) 
vacuum configurations of
the Euclidean theory gauge field, belonging  to the  
distinct topological sectors, are useful, for example, in the 
functional integral quantization of the gauge theory.

\section{Conclusions}\label{concl}
The LF hyperplane is argued to be {\it equally appropriate} as the conventional
one for quantizing field theory. 
The   discussion given above in the {\it front form} 
formulation 
is seen again to be quite  transparent and the physical contents 
following from the {\it quantized} theory 
agree with those known in the conventional {\it instant form} 
treatment. Evidently, they 
should  not depend on whether we employ the  
conventional or the LF coordinates to span the Minkowski space 
and study the temporal evolution of the quantum dynamical system 
in $t$ or $\tau$ respectively. 

We note that in our context the (LF) 
hyperplanes $x^{\pm}=0$ define the characteristic surfaces of 
hyperbolic partial differential equation.    
It is known from their mathematical theory \cite{sne} 
that a solution exists if we specify the initial data 
on both of the hyperplanes. From the present discussion 
and the earlier works \cite{pre, pre1} we conclude that 
it is sufficient in the {\it front form} treatment to choose 
one of the hyperplanes, as proposed by Dirac \cite{dir},  
for canonically quantizing the theory. 
The equal-$\tau$  commutators of the field operators, 
 at a fixed 
initial LF-time, form now  a part of the initial data instead and we
deal with operator differential equations. 
The information on the commutators on the other characteristic 
hyperplane seems to be already contained \cite{pre1} 
in the quantized theory and need not be specified separately. 
As a side comment, the well accepted notion 
that a  classical model field 
theory must be upgraded first through quantization, 
before we confront it  with the 
experimental data, finds here in a sense a theoretical confirmation.

The {\it physical} Hilbert space is obtained in a direct fashion 
in the LF quantized  CSM and SM gauge theories, once 
the constraints are eliminated and the appreciably 
reduced set of independent operators on the LF phase space
identified. The CSM has in it both the massive and the massless 
scalar excitations while only the massive one appears in the 
SM. There are no condensate or $\theta$ vacua in CSM but 
they both have  degenerate vacuum structure. 
In the conventional 
treatment \cite{low} an extended phase space is employed and 
suitable constraints are required to be imposed in order to define the 
{\it physical} Hilbert space which would then lead to the 
description of the physical vacuum state. The existence of one more 
kinematical generator on the LF and  the inherent symmetry in $x^{\pm}$ 
in the quantized theory seem to introduce already 
sufficient number of constraints in the theory leading to great deal  of 
simplifications. Many of the ingredients  like, for example,  
the continuuum normalization 
of the $\theta$-vacua in SM, which needs to be imposed in the conventional 
treatment are already to be found in the {\it front form } quantized theory.   
The functional integral 
method together with the LF quantization may be an efficient tool for
handling the nonperturbative calculations.

A discussion  parallel to 
the one given here can also be made in the {\it front form} theory of
the gauge invariant formulation \cite{abd} of the CSM. 
In an earlier work \cite{pre4}, where  the BRST-BFV functional integral 
quantization was employed, it was demonstrated 
that  this formulation and the gauge noninvariant one in (2) 
in fact lead to the same effective action. 
Also the BRS-BFT  quantization method  proposed \cite{bat}
recently can  be extended to the {\it front form} theory. 
It was applied \cite{pre5} 
to the action (2) and different equivalent effective actions 
obtained for the CSM. 

\baselineskip=16pt

\begin{center}
 {\bf Acknowledgements}
\end{center}

The author would like to thank Werner Israel, 
Silvio Sorella, James Vary, Vitor Oguri, Sebasti\~ao Dias 
and Francisco Caruso  for constructive remarks.  
The hospitality offered at the SLAC Theory Group and  
the financial aid from the CNPq, Brasil, for his 
participation at the  8th Workshop on
 Light-Cone QCD,  Lutsen, MN, 
are gratefully acknowledged.

\end{document}